\documentclass[12pt]{iopart}
 \pdfoutput=1
\usepackage{iopams}
\usepackage{graphicx}
\usepackage{units}
\usepackage{blindtext}
\usepackage{soul}
\usepackage{xcolor}

\usepackage{fakeMnSymbol} 
\begin{document}
\title[Atmospheric continuous-variable quantum communication]{Atmospheric continuous-variable\\quantum communication}
\author{B~Heim$^{1,2,3}$\footnote[1]{Contributed equally to this work}, C~Peuntinger$^{1,2}$\footnotemark[1], N~Killoran$^{4}$, I~Khan$^{1,2}$, C~Wittmann$^{1,2}$, Ch~Marquardt$^{1,2,3}$, and G~Leuchs$^{1,2,3,5}$}
\address{$^1$ Max Planck Institute for the Science of Light, G\"unther-Scharowsky-Str. 1 / Bldg~24, 91058 Erlangen, Germany}
\address{$^2$ Institute of Optics, Information and Photonics, Friedrich-Alexander-Universit\"at Erlangen-N\"urnberg~(FAU), Staudtstr. 7/B2, 91058 Erlangen, Germany}
\address{$^3$ Erlangen Graduate School in Advanced Optical Technologies (SAOT), FAU, Paul-Gordan-Str. 6, 91052 Erlangen, Germany}
\address{$^4$ Institute of Theoretical Physics, Ulm University, Albert-Einstein-Allee 11, 89069 Ulm, Germany}
\address{$^5$ Department of Physics, University of Ottawa, Ottawa, Ont. K1N 6N5, Canada}

\ead{bettina.heim@mpl.mpg.de}
\begin{abstract}
We present a quantum communication experiment conducted over a point-to-point free-space link of \unit[1.6]{km} in urban conditions.
We study atmospheric influences on the capability of the link to act as a continuous-variable (CV) quantum channel.
Continuous polarization states (that contain the signal encoding as well as a local oscillator in the same spatial mode) are prepared and sent over the link in a
polarization multiplexed setting. Both signal and local oscillator undergo the same atmospheric fluctuations. These are intrinsically auto-compensated which removes detrimental influences on the interferometric visibility. At the receiver, we measure the Q-function and interpret the data using the framework of effective entanglement.
We compare different state amplitudes and alphabets (two-state and four-state) and determine their optimal working points with respect to the distributed effective entanglement. Based on the high entanglement transmission rates achieved, our system indicates the high potential of atmospheric links in the field of CV QKD.
\end{abstract}
\maketitle
\section{Introduction}
Quantum communication refers to the distribution of quantum states between two parties via a quantum channel. With regard to this, it is crucial that this quantum channel preserves the quantum properties of the distributed states. The most common channel implementation are optical fibers and the free space. The latter offers great flexibility in terms of infrastructure establishment and  links to moving objects are also feasible, see e.g.~\cite{nauerth13}. For a review of representative free-space quantum communication experiments see~\cite{meyers10}.

Quantum key distribution (QKD)~\cite{gisin02,scarani08} is probably the most practical branch of quantum communication and concerns the establishment of a secret key jointly between two legitimate parties, Alice and Bob. As the security is based on the laws of quantum mechanics, in principle information theoretic-security can be achieved~\cite{scarani08}.

Free-space QKD over a real atmospheric channel was first demonstrated in 1996~\cite{jacobs96}. Since then, a number of prepare-and-measure as well as entanglement based schemes have been implemented in free space; the longest distance so far achieved on Earth is \unit[144]{km}~\cite{schmitt-manderbach07,ursin07a}.
Nowadays, even quantum communication between earth and space is being conceived~\cite{rarity02,aspelmeyer03,villoresi08,perdiguesarmengol08,bonato09,meyer-scott11, bourgoin13}. All of the systems so far referred to here have one major aspect in common: they are based on discrete quantum variables and use single photon threshold ("click") detectors, which involves spatial, spectral and/or temporal filtering in order to reduce background noise (see e.g.~\cite{hughes00,buttler00} for the first point-to-point demonstrations of free-space QKD in daylight).

As already shown by Bennett~\cite{bennett92}, any two non-orthogonal quantum states suffice to ensure secure key distribution. This paved the way for continuous-variable protocols (for a review see~\cite{andersen09,weedbrook12}), based on a different approach: performing homodyne measurements on weak coherent states with the help of a bright local oscillator (LO). Generally, homodyne detectors offer immunity to stray light and the PIN photodiodes used therin own higher quantum efficiencies than avalanche photodiodes. Initially, a discrete modulation of Gaussian states was proposed ~\cite{ralph99,hillery00,reid00}, followed by Gaussian modulation~\cite{cerf01, grosshans03} shortly thereafter. 

The homodyning technique is well established in classical optical communication including satellite based links~\cite{heine11}. There, the signal states are at high intensities, their overlap is negligible, and generating a LO locally at the receiver is appropriate.
CV QKD uses strongly overlapping and thus almost indistinguishable states, which requires high detection efficiencies as well as a high interferometric visibility. We fulfilled this demand by means of a specially developed protocol using the polarization degree of freedom to multiplex signal and LO in the same spatial channel mode already at the sender~\cite{lorenz04,lorenz06}. A successful pilot project proved the feasibility of CV QKD via a \unit[100]{m} free-space link on the roof of our institute~\cite{elser09,heim10}.
We now succeeded to established a point-to-point free-space link connecting two buildings at a distance of \unit[1.6]{km} in an urban environment within the city of Erlangen. A two-dimensional modulation of the signal states and their simultaneous detection along two conjugate quadratures (as initially shown in~\cite{lorenz04} and~\cite{lance05}) was implemented, using the aforementioned polarization multiplexing. For the moment, we focus on the discrete modulation of two or four signal states, but in principle, a variety of alphabets such as Gaussian modulation~\cite{cerf01,grosshans03} or ring-type alphabets~\cite{sych10} can be realized with our system. Note that due to the inherent auto-compensation of phase fluctuations, a freely drifting phase, such as might be the case with fiber channels~\cite{wittmann10}, can be avoided.\\
Effective entanglement (EE)~\cite{rigas06,haeseler08} is a powerful measure in describing the quantum correlations and their preservation during channel transmission  and a necessary precondition for QKD.
We verify and quantify EE in terms of minimal negativity~\cite{killoran11,khan13} according to the method introduced in~\cite{rigas06,haeseler08,killoran10,killoran11}. In our free-space scenario, special consideration has to be given to atmospheric turbulence that not only affects the signal states but also leads to fluctuations of the LO and to undesired classical excess noise in general. Such effects on quantum continuous-variable states in a turbulent atmosphere have been studied recently from a theoretical~\cite{semenov09,usenko12,semenov12} as well as an experimental~\cite{elser09,heim10,peuntinger14} point of view. Furthermore, there have also been ideas suggesting to exploit the fading channel properties in terms of improved quantum state propagation~\cite{heersink06,dong08,schnabel08,wittmann08,semenov09,usenko12,erven12,capraro12}.

\subsection{Stokes operators}
In quantum mechanics, polarization can be described by the quantum Stokes operators, which are the quantum counterparts of the classical Stokes parameters~\cite{stokes1852}. The Stokes operators are introduced and defined, for example, in~\cite{korolkova02} and read as 
\begin{eqnarray}
\hat{S}_{0} &= \hat{a}^{\dag}_{\mathrm{H}} \hat{a}_{\mathrm{H}} + \hat{a}^{\dag}_{\mathrm{V}} \hat{a}_{\mathrm{V}}&~\mathrm{\it(total~intensity)},\\
\hat{S}_{1} &= \hat{a}^{\dag}_{\mathrm{H}} \hat{a}_{\mathrm{H}} - \hat{a}^{\dag}_{\mathrm{V}} \hat{a}_{\mathrm{V}} &~(\leftrightarrow - \updownarrow),\\
\hat{S}_{2} &= \hat{a}^{\dag}_{\mathrm{H}} \hat{a}_{\mathrm{V}} + \hat{a}^{\dag}_{\mathrm{V}} \hat{a}_{\mathrm{H}} &~ (\MNSneswarrow- \MNSnwsearrow), \\
\hat{S}_{3} &= i(\hat{a}^{\dag}_{\mathrm{V}} \hat{a}_{\mathrm{H}} - \hat{a}^{\dag}_{\mathrm{H}} \hat{a}_{\mathrm{V}}) &~ (\MNSlcirclearrowright-\MNSrcirclearrowleft)
\end{eqnarray}
in terms of the creation and annihilation operators $\hat{a}^{\dagger}$ and $\hat{a}$ of the respective polarization modes.
The arrows in brackets display the operational definitions of the Stokes parameters as intensity differences of polarization types.
In our experiment, where the LO is circularly polarized and the signal states are measured by a homodyne detection of the $S_{1/2}$-Stokes operators (see Figure~\ref{fig:setup}), this results in the uncertainty relation: $\textrm{Var} ( \hat{S}_1 ) \cdot \textrm{Var} ( \hat{S}_2 ) \geq |\langle\hat{S}_3\rangle|^{2}$.
In the case of coherent states, the equality holds and the variances of $\hat{S}_1$ and $\hat{S}_2$ are equal. Polarization excess noise results in increased variances.
The co-propagation of signal and LO ensures a perfect spatial interference between the two, enabling high detection efficiency without any additional interference stabilization, as channel induced phase fluctuations are auto-compensated. For our free-space system, the co-propagation has further advantageous side effects: the LO acts as a spatial and spectral filter, such that only those photons that are mode-matched with it will result in a significant detector signal. In contrast to what is the case in single photon experiments, here there is no need for spatial or spectral filtering, and background light (that is not mode matched with the LO) does not disturb the measurement.

\section{Experimental Setup}
\begin{figure}
	\centering
		\includegraphics[width=0.9\textwidth]{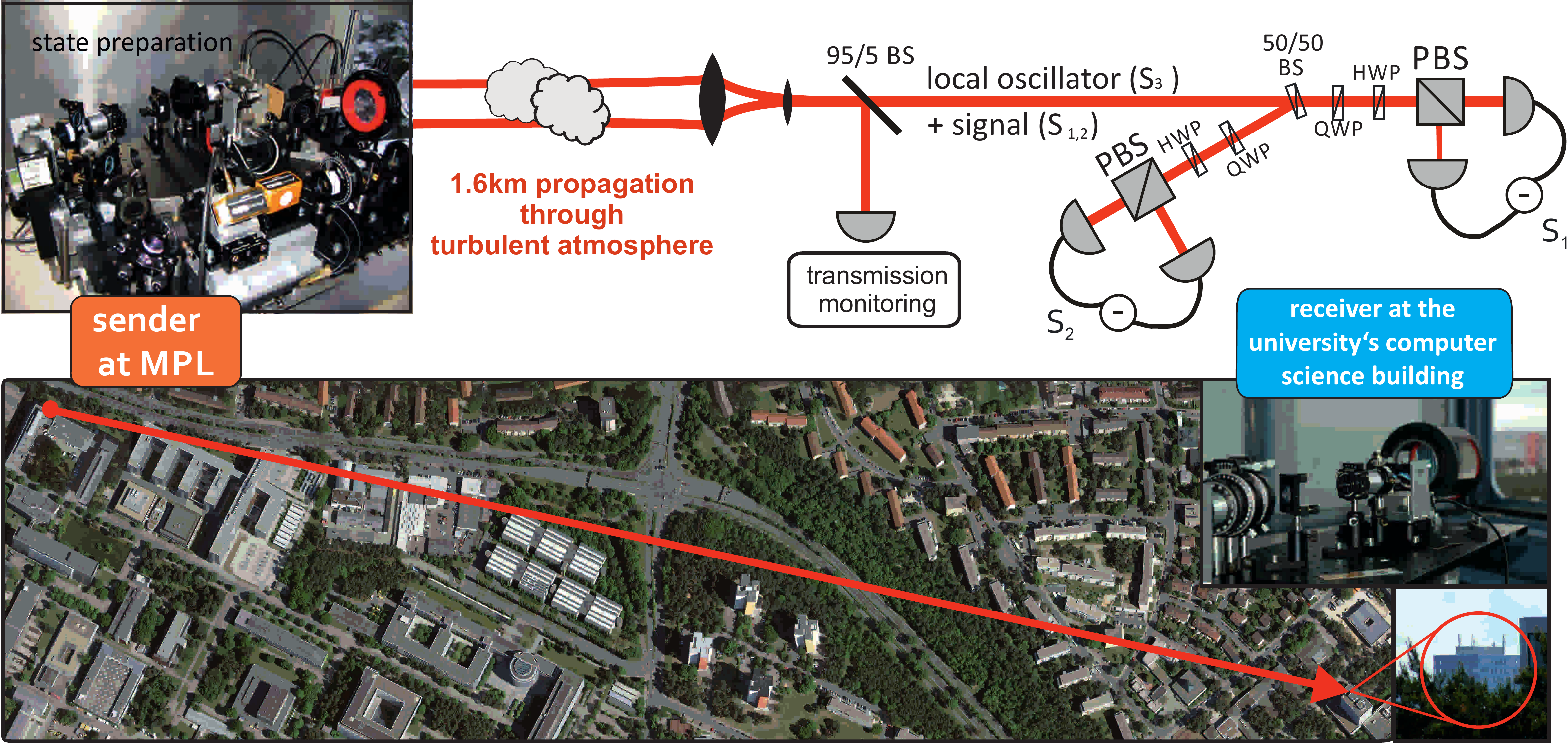}
	\caption{Schematic and aerial view of the free-space link of \unit[1.6]{km} within the city of Erlangen. After state preparation at the sender, both signal and local oscillator are spatially expanded and sent to the receiver, polarization multiplexed and in the same spatial mode. A telescope with a front aperture of \unit[150]{mm} again reduces the beam diameter. After splitting off a small part for transmission monitoring, the signal states are detected by a double homodyne measurement, realized as a simultaneous Stokes measurements of the $S_{1/2}$ parameters.
HWP: half-wave plate; QWP: quarter-wave plate, (P)BS: (polarizing) beam splitter.
Picture rights: Orthophoto \copyright~Bayerische Vermessungsverwaltung 2013}
	\label{fig:setup}
\end{figure}

The optical setup of our QKD system is shown in Figure~\ref{fig:setup}. At the sender, a grating stabilized continuous wave diode laser (Toptica DLL 100) at a wavelength of \unit[809]{nm} is spatially mode cleaned and guided to an alignable breadboard (see upper left part in Figure~\ref{fig:setup}) via a single mode polarization maintaining fiber. A small portion is then split off to enable monitoring of the shot-noise limitation of the laser output states in a self-homodyning setup. With the remaining portion, we prepare the circularly polarized LO using a sequence of two polarizing beam splitters (PBS), a half wave plate (HWP) and a quarter wave plate (QWP). Two electro-optical modulators (Thorlabs, EO-AM-NR-C1, \unit[600-900]{nm}, bandwidth \unit[100]{MHz}), that sandwich a HWP for $45^{\circ}$ polarization rotation, are voltage driven by two synchronized arbitrary waveform generators (Agilent 33250A) and serve for the generation of signal states using the Pockels effect. At a repetition rate of \unit[3.05]{MHz}, Gaussian shaped voltage pulses in the order of only few $\mathrm{mV}$ vary the crystal's birefringence. They cause a time dependent phase shift, diminishing the amount of circular (LO) polarization and introducing weak signal components. The signal energy is taken from the LO, which, due to the weak modulation, remains essentially unaffected. Each signal pulse is followed by a zero voltage period of the same length, acting as the vacuum reference. After 263 signal pulses, the modulation voltage is increased  
for exactly one pulse that later serves as the trigger for a clock recovery in the receiving process.
The state generation process is verified by a double homodyne detection scheme realized as a simultaneous measurement of the $S_{1/2}$-Stokes parameters. The two outputs of a non polarizing, symmetric beam splitter are equipped with a combination of a quarter and a half wave plate (to compensate for polarization offsets and to adjust the $S_{1/2}$ basis), a polarizing beam splitter and a homemade detector, respectively.
The required linearity of the detection system is verified by an attenuation measurement of the unmodulated LO. The detectors measure the electronic difference current of two photo diodes (the overall detection efficiency including optical losses and the diodes' quantum efficiencies is $0.84\pm 0.02$.) and amplify the translated voltage. The signal obtained is then electronically high-pass filtered (Minicircuits BLK-89-S+, \unit[100]{kHz})\footnote{Note that we verified that the influence of this filter on the signal pulses at a repetition rate of \unit[3.05]{MHz} is negligible.} to compensate for long term drifts, and measured by an oscilloscope at a sampling rate of \unit[250]{MHz}. One time slot, containing signal and vacuum reference, consists of 82 samples; the signal state - or corresponding vacuum - is determined by integrating over 41 samples. To prevent any influence of the trigger pulse on the quantum signals, we disregard the first 92 out of 264 slots, leading to an effective state modulation rate of \unit[2.22]{MHz}.

\begin{figure}
	\centering
		\includegraphics[width=0.9\textwidth]{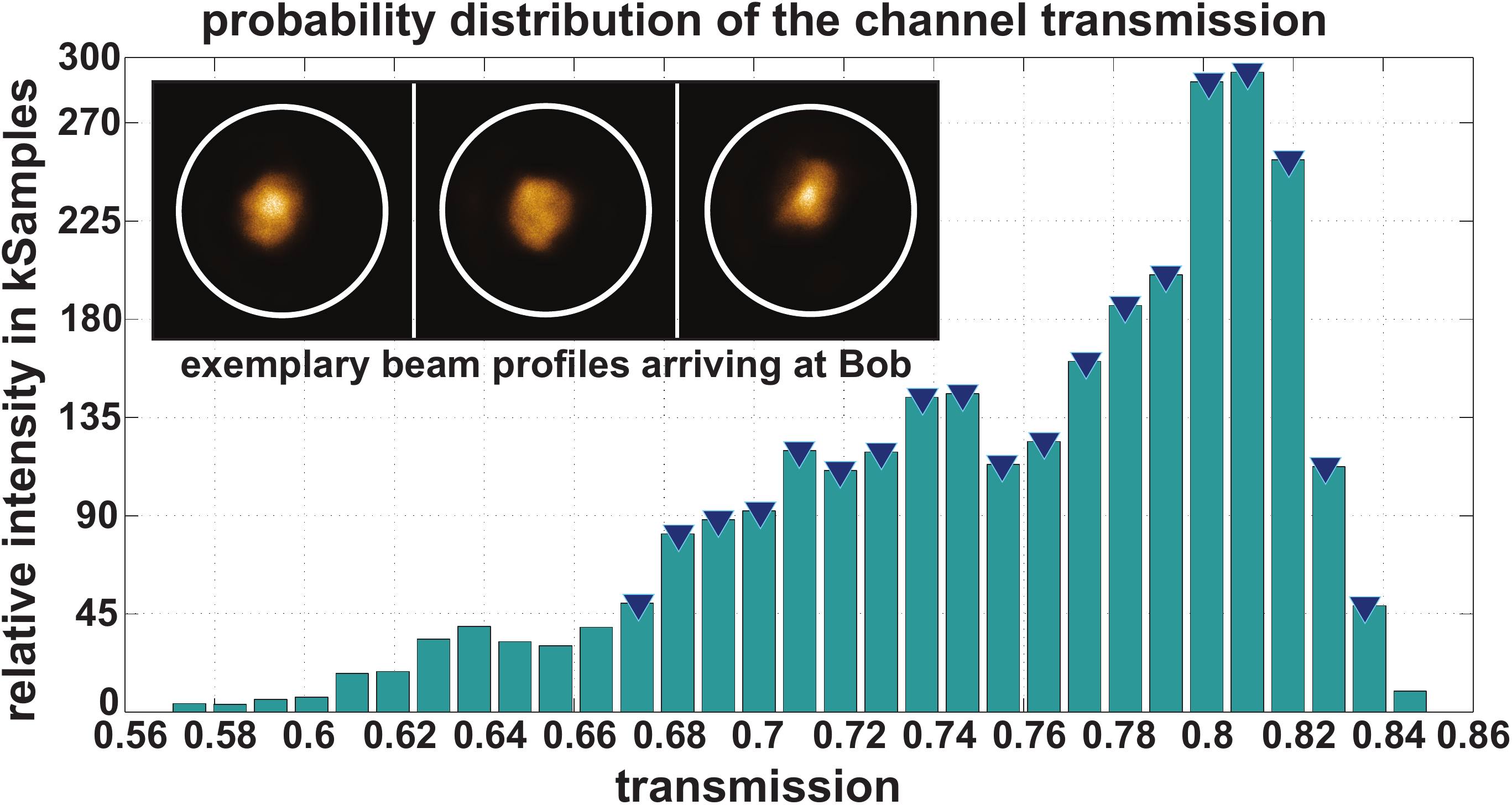}
	\caption{Probability distribution of the fluctuating channel transmission $T$, where $T_{mean}$ is~$\approx 76.1\%$. The monitoring measurement is performed by permanently tapping off $<4\%$ of the arriving beam. The width of one of the 35 bins ($\Delta T$) is 0.9\%. The triangles indicate those bins that had sufficiently large sample sizes and were thus used for further processing. They contain \unit[92]{\%} of events.
 The inset shows three example shots of the spatial beam profile, taken at an exposure time of \unit[0.64]{ms}, while the circle labels the \unit[150]{mm} receiver aperture.}
	\label{fig:Bob_pdtc}
\end{figure}

After the state preparation, the signal and LO beam is spatially expanded to $\unit[4]{cm}$ in diameter and sent to the receiver through the \unit[1.6]{km} free-space link. Atmospheric turbulence causes the beam to diverge more than expected by the diffraction limit, leading to an increased beam width after channel propagation. This effect can be attributed to both, wavefront distortions and "beam wandering" or spatial jitter. The latter is visualized in the inset in Figure~\ref{fig:Bob_pdtc} showing three examples of shots of the spatial beam profile arriving at Bob.
At the receiver, a telescope with a front aperture of \unit[150]{mm} is needed to capture most of the arriving beam intensity.
After reduction of the beam diameter and splitting off a small part ($<4\%$) for transmission and LO monitoring, the signal states are detected in the same way as described above for Alice (see Figure~\ref{fig:setup}).
\\
The channel induces intensity fluctuations, Figure \ref{fig:Bob_pdtc} shows the statistical distribution of the channel transmission $T$, monitored in a direct detection simultaneously with the quantum signal.  
The width of one of the 35 bins ($\Delta T$) is 0.9\%, $T_{mean}$ is~$\approx 76.1\%$. The triangles indicate those bins that had sufficiently large sample sizes and were thus used for further processing. They contain $\unit[92]{\%}$ of events.
These measurements were made on a rather clear and calm night (September~$4^{th}$ 2013). We want to point out that daylight operation of our homodyne system is easily feasible without any spectral or spatial filters. However, we chose to measure at night, as atmospheric turbulence is increased during the day, which would directly lead to a broader transmission distribution at a lower mean value.\\

The linearity of Bob's detection setup was equally well verified by an attenuation measurement of the unmodulated LO at intensities down to zero. Here, the overall detection efficiency (not including the monitoring split-off) is $0.83\pm 0.02$. For a typical received LO power of \unit[14]{mW}, the electronic noise is more than $\unit[10]{dB}$ below the shot-noise level. 
The fading channel, however, has further impact on Bob's detection. Most important is the fluctuation of the shot-noise level, that is given by the unmodulated LO. Hence the signal modulation scheme is performed such that each signal slot is followed directly by its corresponding vacuum reference. 
We apply again a high-pass filter to avoid slow drifts in the offset of the signal, as the dynamic range of the oscilloscope's A/D conversion is limited.

\section{Results}
\begin{figure}
	\centering
		\includegraphics[width=0.99\textwidth]{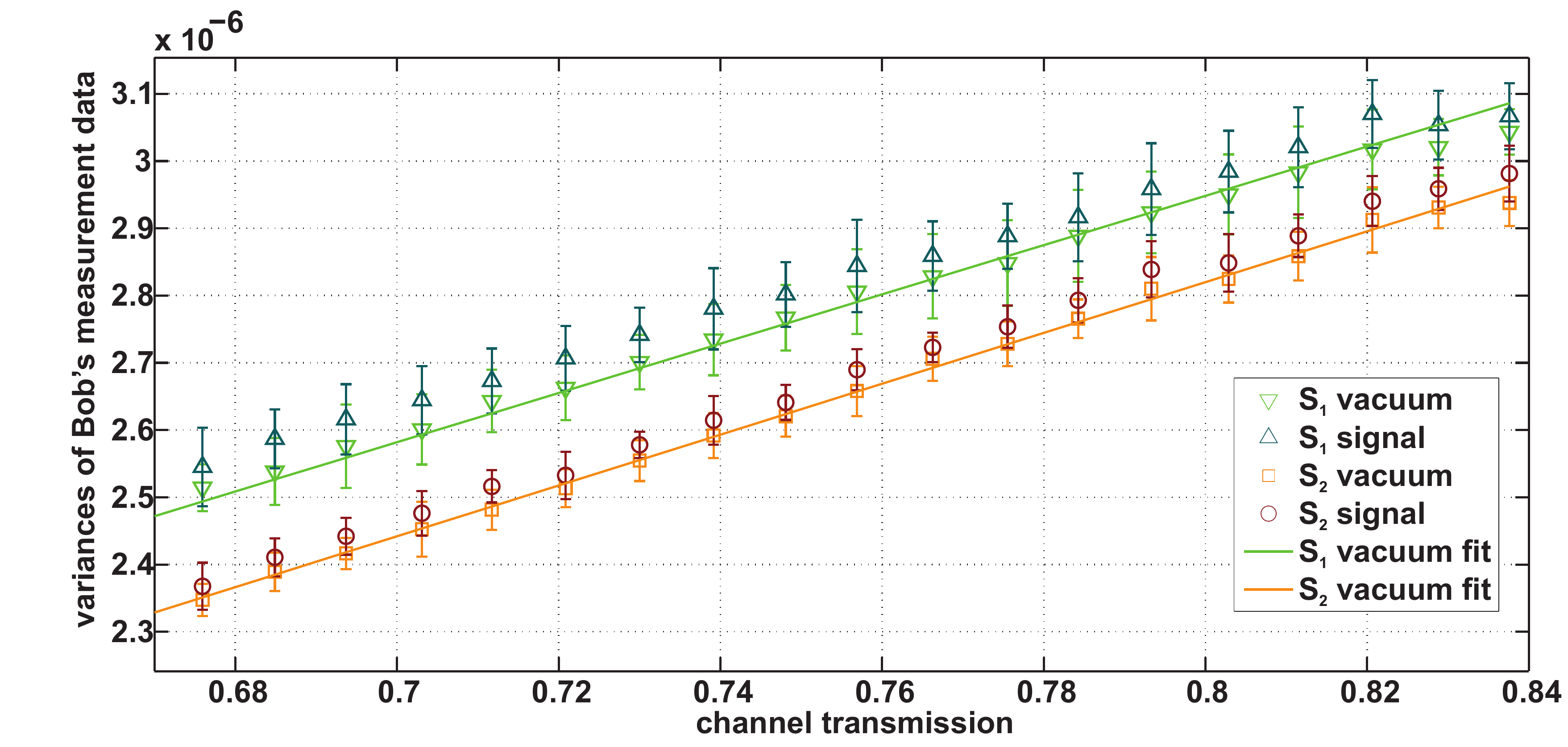}
	\caption{Variances calculated of 15000 signal states and corresponding vacua are plotted against the respective channel transmission for each transmission bin. Only those bins containing statistically sufficient numbers of samples (as indicated by the triangles in Figure \ref{fig:Bob_pdtc}) are considered. We averaged over four states, measured simultaneously along the $S_{1}$ and $S_{2}$ direction. The error is estimated conservatively by one standard deviation. The measurements of the vacuum are consistent with a linear fit, as theoretically expected.}
	\label{fig:Bob_var}
\end{figure}

We use the classical transmission value to sort the collected data according to the prevailing transmission bins~(see~\cite{peuntinger14,usenko12}). This technique allows us to directly compare the measured Stokes parameter variances to the corresponding shot-noise variance.
For an alphabet of four states distributed symmetrically 
($\left | + \alpha \right\rangle$, $\left | + i\alpha \right\rangle$, $\left | - \alpha \right\rangle$,  $\left | - i\alpha \right\rangle$), the variances of 15000 states and corresponding vacua are calculated for each individual transmission bin. We discard the bins that contain too few samples; the remaining data have small statistical errors. The results are shown in Figure~\ref{fig:Bob_var}, averaged over four states and individually for the $S_{1}$/$S_{2}$ direction. The error is estimated by one standard deviation, the expected linearity of the vacuum states is shown by the linear fit.
 The (mean) excess noise of 0.01 shot-noise units (SNU) of the signal states compared to vacuum is mainly of technical origin, and is already introduced in Alice's signal generation step. This is due to the stronger fluctuating voltage pulses applied to the EOMs compared to the unmodulated vacuum slots.
Note that the signal amplitudes at Alice have been chosen near 1, as this was determined as an optimal value for the four state alphabet in terms of negativity guaranteeing the maximal possible rate of effective entanglement. This will be explained in more detail in the next section.

\begin{figure}
	\centering
		\includegraphics[width=0.9\textwidth]{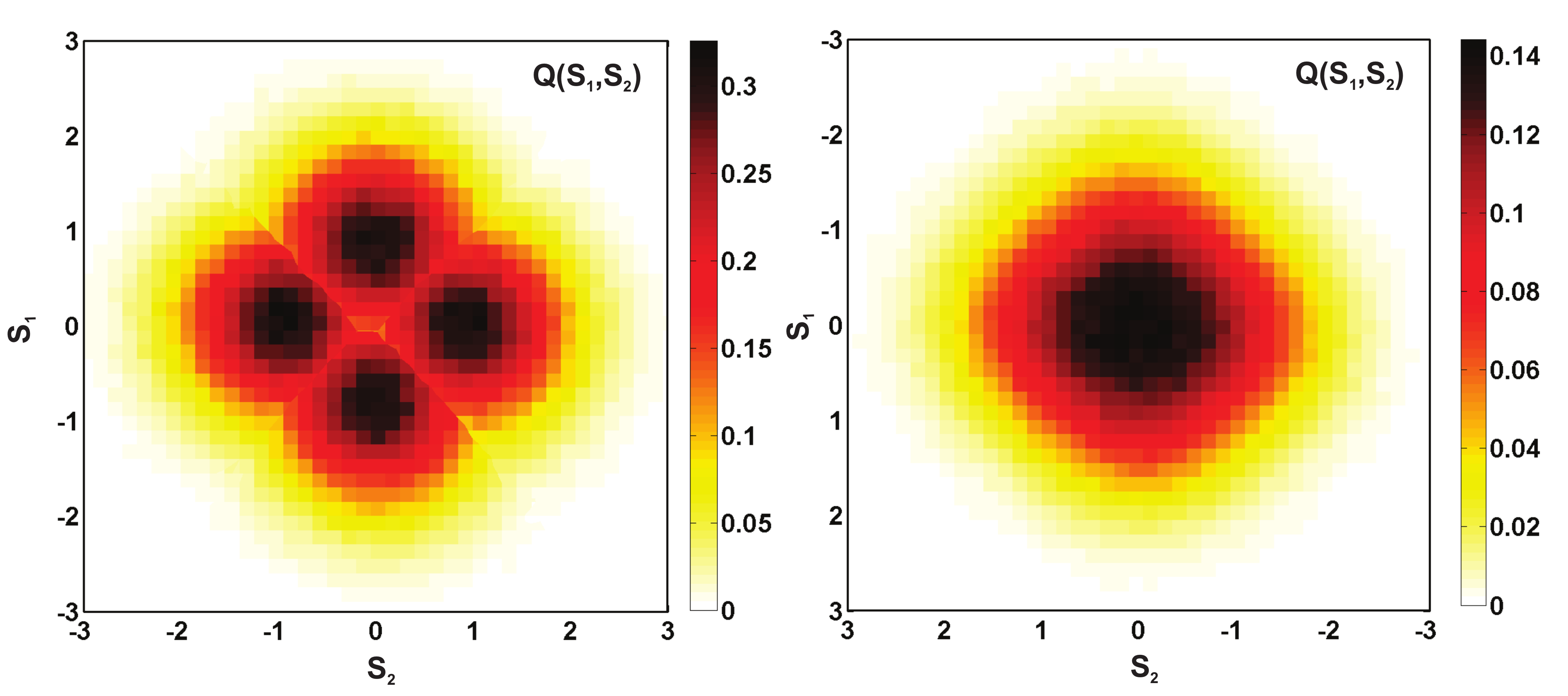}
	\caption{Q-function of four states (each of them containing 270000 samples) for the most frequent bin at 81\% transmission. The signal amplitudes are around 0.9, the bin size is 0.15, as indicated by the contour plot edges. The left hand side shows each state plotted individually as well as the combined Q-function of the mixed state (right). The individual states almost reach the maximal height of the Q-function of $\frac{1}{\pi}$ related to a pure state. The peak height of the mixed state is distinctly less ($\approx~0.14$).}
	
	\label{fig:Bob_Qfct}
\end{figure}

For the largest occupied transmission bin, the Q-function~\cite{stenholm92} of four signal states received at Bob was measured; each signal state dataset contained 270000 samples. Figure~\ref{fig:Bob_Qfct} shows on the left hand side each state plotted individually as well as the combined Q-function of the mixed state (right). 
The height of the Q-function is an indicator of mixedness of the depicted state~\cite{lorenz06}. 
The maximal value of $\frac{1}{\pi}$ can be achieve by a pure state, as e.g. the vacuum state (not shown here). The individual states (left) are almost pure, the peak height of the mixed state is distinctly less ($\approx 0.14$), as theoretically expected.  

The signal amplitudes, which have been attenuated by the channel transmission of $\approx 81.2\%$, are: 0.88, -0.87 ($S_{1}$ observable) and  0.92, -0.92 ($S_{2}$ observable). The slight asymmetry is already introduced at Alice, mainly by the fact that the same voltages are applied to both EOMs, the transmission of which is $\approx 95\%$. As the first EOM modulates along the $S_{1}$ axis, these signals are slightly attenuated.

\begin{figure}
	\centering
		\includegraphics[width=0.9\textwidth]{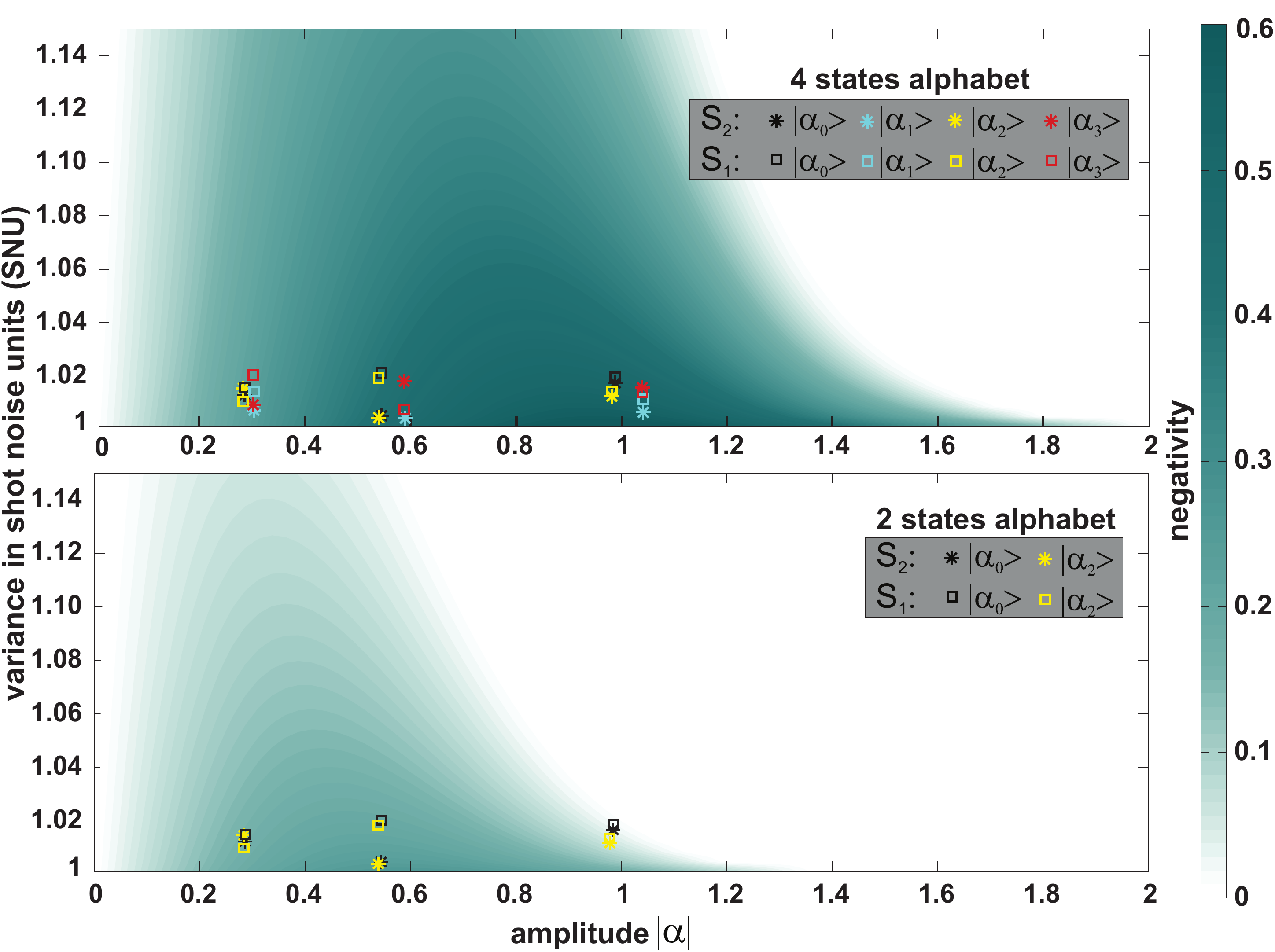}
	\caption{Comparison of effective entanglement quantification in a two- vs. a four-state alphabet.
 The negativity bounds in the shaded region were calculated theoretically by assuming all variances
equal and thus only serve as a reference for the slightly varying experimental values (inset).
For a given transmission of $\approx 63\%$ (referring to its mean value, including detector efficiency), the four-state alphabet clearly offers larger values of negativity and tolerates significantly more excess noise. The optimal amplitudes also strongly depend on the choice of alphabets.}
	\label{fig:Bob_2vs4states}
\end{figure}

\textbf{Effective entanglement verification and quantification}\\
The use of entanglement verification describing correlations of a bipartite state shared between Alice and Bob as a first necessary precondition has been proposed in the context of QKD protocols~\cite{curty04,rigas06}. This is especially useful in the absence of a full security proof considering realistic amounts of excess noise as well as finite size effects for discrete modulation CV QKD protocols. (Still, promising advances in this direction have been made within the last years~\cite{zhao09,leverrier09er11,leverrier11}). The method was extended to the Stokes operators (\cite{haeseler08}, experimental realization in~\cite{lorenz06,wittmann08}), where monitoring of the LO can help identify the impact of an adversary.
Following~\cite{killoran11,khan13}, we use the negativity~\cite{zyczkowsky98,lee00,vidal02} as an entanglement measure that can be minimized over all bipartite states consistent with the information available. By means of semi-definite programming~\cite{mosekyalmip}, we find optimal working points  with respect to the initial state overlap and measured excess noise for our continuous-variable system. Formally, this treatment applies to quadrature operators, not Stokes operators. As in~\cite{khan13}, we obtain quadrature measurements from Stokes measurements by considering a specific trusted form for the local oscillator. As shown in Figure~\ref{fig:Bob_2vs4states}, we compare two alphabets using discrete modulation of two versus four signal states. The four-state alphabet clearly outperforms the two-state counterpart in terms of acceptable excess noise and higher negativity bounds. The optimal amplitudes also depend greatly on the choice of alphabets. 
For these plots, the transmission was assumed to be fixed at $\unit[63]{\%}$ (referring to its mean value, including detector losses).

\begin{figure}
	\centering
		\includegraphics[width=0.9\textwidth]{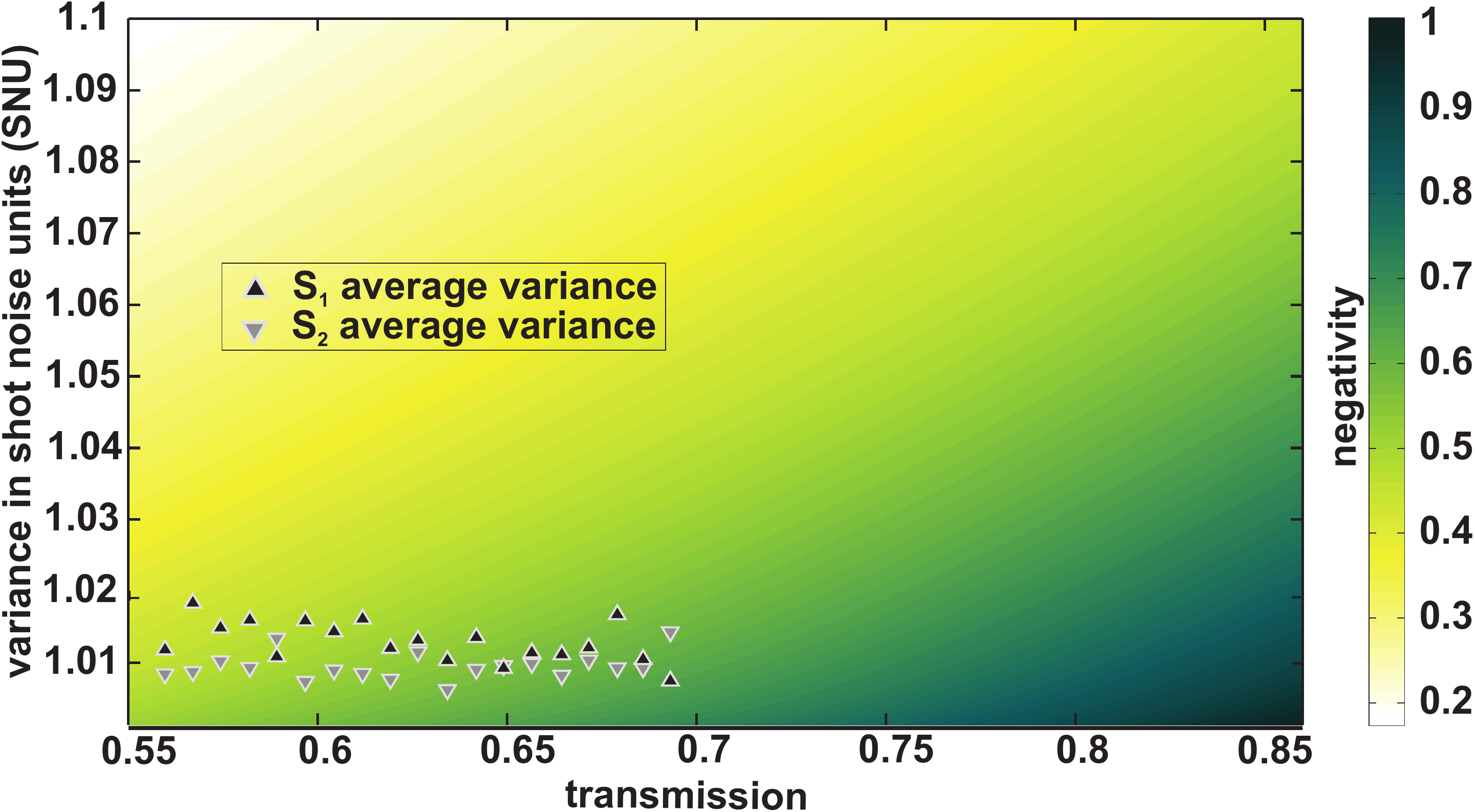}
	\caption{Effective entanglement minimization performed individually for each transmission sub-channel, for the four-state alphabet at a fixed and optimal amplitude near 1.	The negativity bounds in the shaded region were calculated theoretically by assuming all variances equal and thus only serve as a reference for the slightly varying experimental values (see inserted mean variances). At a channel transmission close to 100\% (or 83\% with detection efficiency considered), the entanglement transfer would be optimal.}
	\label{fig:EE_Tvar}
\end{figure}

\begin{figure}
	\centering
		\includegraphics[width=0.9\textwidth]{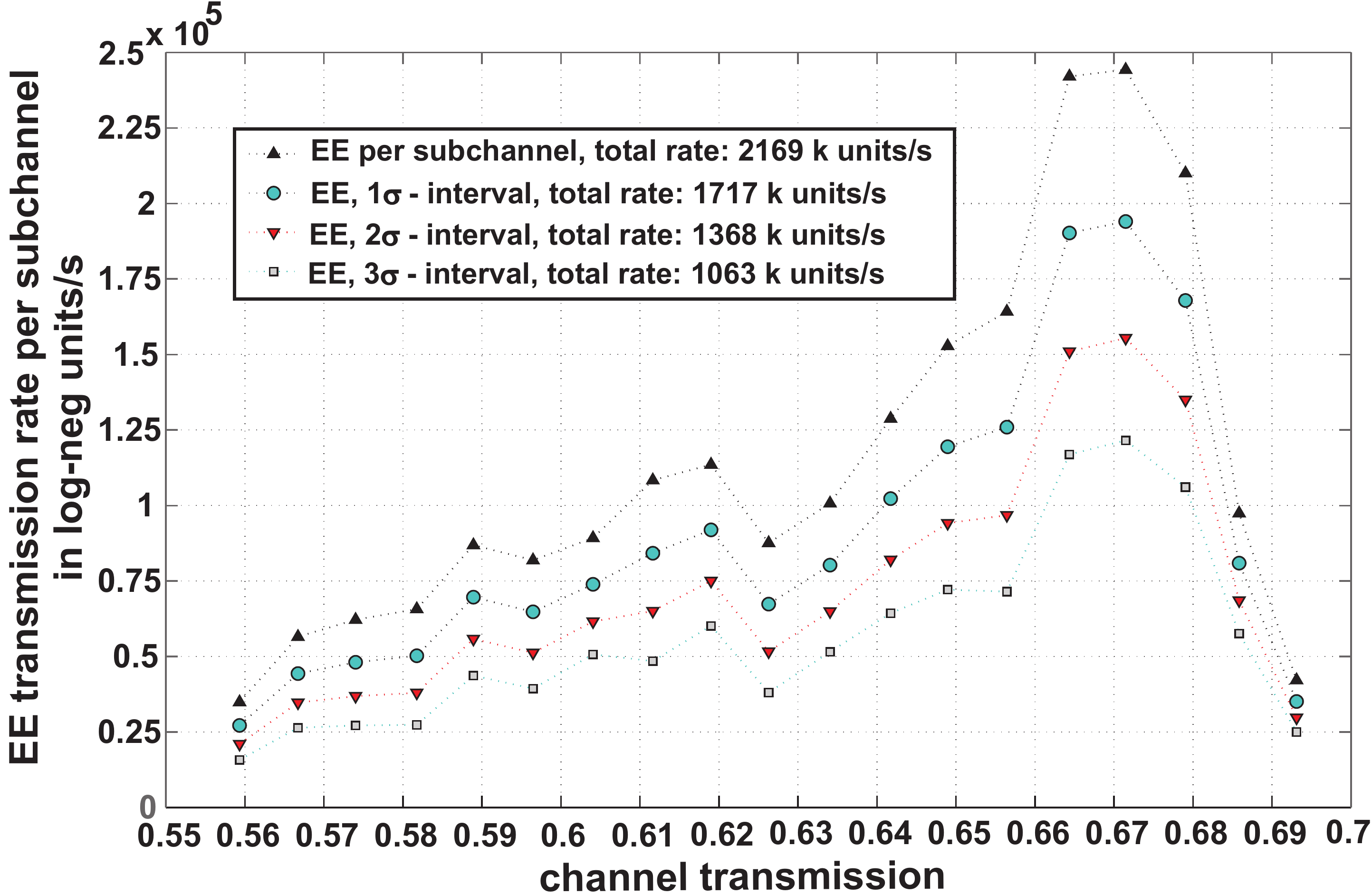}
	\caption{Minimal rate of distributed entanglement per subchannel. By using the logarithmic negativity~\cite{plenio05}, we can sum over all sub-channels, weighted according to their frequency of occurance. This leads to an overall entanglement transfer rate of \unit[2.17] {M log-neg units/s}. By expanding the constraint regions in the minimization process, we gain $1\sigma$, $2\sigma$ and $3\sigma$ intervals, with respectively lower average entanglement distribution rates.
 }
	\label{fig:EE_rates}
\end{figure}

Unlike the fixed channel transmission of the experimental fiber based setup in~\cite{khan13}, we have to cope with the naturally fluctuating transmission.
For the four-state alphabet, we now perform the minimization for each of the transmission sub-channels individually, at a fixed and optimal amplitude near 1. The result is shown in Figure~\ref{fig:EE_Tvar}. At a channel transmission close to 100\% (or 83\% with detection efficiency considered), obviously, the entanglement transfer would be optimal.
For the estimation of an overall entanglement transfer rate, we use the concept of logarithmic negativity~\cite{plenio05}, as it is additive. Figure~\ref{fig:EE_rates} shows the individual rates for each sub-channel. We sum over all of them, weighted according to their frequency of occurance. This leads to an average minimal entanglement transfer rate of \unit[2.17] {M log-neg units/s}. We also performed the entanglement minimization while taking into account experimental error bars. Using standard error propagation techniques, these measurement uncertainties lead to looser constraint sets for the minimization (this is the same approach taken in~\cite{khan13}). We optimized at 1$\sigma$, $2\sigma$ and $3\sigma$ confidence levels, each giving progressively lower overall entanglement distribution rates.

\section{Conclusion}
We experimentally demonstrated the preservation of quantum properties through the verification of EE for an atmospheric point-to-point link for CV quantum communication in an urban environment.  We compare a two-state and four-state alphabet of discretely modulated signal states. The four-state alphabet clearly outperforms the two-states in terms of potentially achievable negativity and tolerable excess noise. Thus we determined the optimal working points for the latter and calculated the rate of distributed entanglement, also taking into account the channel induced drifting shot-noise variance.
Note that even though we focused on discrete modulation so far, in principle Gaussian modulation~\cite{cerf01,grosshans03} can also be easily realized with our system.
The shown results clearly indicate the strong potential of free-space channels for CV QKD systems. In addition, they pave the way for further free-space CV quantum communication experiments such as CV entanglement-based QKD~(\cite{madsen12} and see~\cite{peuntinger14} for a demonstration of the free-space distribution of squeezed states) or the use of CV quantum states for atmospheric sensing.

\ack
The project was supported under FP7 FET Proactive by the integrated project Q-Essence and CHIST-ERA (Hipercom). The authors acknowledge the support of the Erlangen Graduate School in Advanced Optical Technologies (SAOT) by the German Research Foundation (DFG) within the framework of the German excellence initiative.\\
The authors thank Norbert L{\"u}tkenhaus for valuable comments and our colleagues at the FAU computer science building for their kind support and for hosting our receiver setup.

\section*{References}
\bibliographystyle{iopart-num}

\begin{thebibliography}{60}
\bibitem{nauerth13} Nauerth~S, Moll F, Rau M, Fuchs C, Horwath J, Frick S and Weinfurter H 2013  \textit{Nat. Photon.} \textbf{7} 382
\bibitem{meyers10}
Tunick A, Moore T, Deacon K and Meyers R 2010 \textit{Quantum Communications and Quantum Imaging VIII, Proc. of SPIE} Vol. \textbf{7815} 781512
\bibitem{gisin02}
Gisin N, Ribordy G, Tittel W and Zbinden H 2002 \textit{Rev. Mod. Phys.} \textbf{74}, 145
\bibitem{scarani08}
Scarani V, {Bechmann-Pasquinucci} H, Cerf N J , Du\v{s}ek M, L\"{u}tkenhaus N and Peev M 2009 \textit{Rev. Mod. Phys.} \textbf{81} 1301
\bibitem{jacobs96}
Jacobs B C and Franson J D 1996 \textit{Opt. Lett.} \textbf{21} 1854
\bibitem{schmitt-manderbach07}
 {Schmitt-Manderbach} T \textit{et al.}  2007 \textit{Phys. Rev. Lett.} \textbf{98} 010504
\bibitem{ursin07a}
 Ursin R \textit{et al.}  2007 \textit{Nat. Phys.} \textbf{3} 481
\bibitem{rarity02} Rarity J G,  Tapster P R, Gorman P M  and Knight P 2002 \textit{New J. Phys.} \textbf{4} 82
\bibitem{aspelmeyer03} Aspelmeyer M, Jennewein T, Pfennigbauer M, Leeb W and Zeilinger A 2003 \textit{IEEE J. Sel. Top. Quantum Electron.} \textbf{9} 1541
\bibitem{villoresi08}
Villoresi P \textit{et al.} 2008 \textit{New J. Phys.} \textbf{10} 033038
\bibitem{perdiguesarmengol08}
 {Perdigues Armengol}~J~M \textit{et al.} 2008 \textit{Acta Astronaut.} \textbf{63} 165
\bibitem{bonato09} Bonato C, Tomaello A, Da Deppo V, Naletto G and Villoresi P 2009 \textit{New J. Phys.} {\bf 11} 045017
\bibitem{meyer-scott11} Meyer-Scott E, Yan Z, MacDonald A, Bourgoin J-P, H\"ubel H and Jennewein T 2011 \textit{Phys. Rev. A} {\bf 84} 062326
\bibitem{bourgoin13} Bourgoin~J-P \textit{et al.} 2013 
\textit{New J. Phys.} \textbf{15} 023006
\bibitem{hughes00} 
Hughes~R~J, Buttler~W~T, Kwiat~P~G, Lamoreaux~S~K, Morgan~G~L, Nordholt~J~E and Peterson~C~G 2000 \textit{J. Mod. Opt. }\textbf{47} 549
\bibitem{buttler00}
Buttler~W~T, Hughes R~J, Lamoreaux~S~K,Morgan~G~L, Nordholt~J~E and Peterson~C~G 2000
\textit{Phys. Rev. Lett.} \textbf{84} 5652
\bibitem{bennett92}
Bennett C H 1992 \textit{Phys. Rev. Lett.} \textbf{68} 3121
\bibitem{andersen09}
Andersen U~L, Leuchs G and Silberhorn C~ 2010 \textit{ Laser \& Photon. Rev.} \textbf{4} 337
\bibitem{weedbrook12}
Weedbrook C,  Pirandola S, Garc\'{i}a-Patr\'{o}n R, Cerf N J, Ralph T C, Shapiro J H and Lloyd S 2012 \textit{\RMP} \textbf{84} 2 621
\bibitem{ralph99}
 Ralph T C 1999 \textit{Phys. Rev. A} \textbf{61} 010303
\bibitem{hillery00}
Hillery M 2000 \textit{Phys. Rev. A} \textbf{61} 022309
\bibitem{reid00}
Reid M D  2000 \textit{Phys. Rev. A} \textbf{62} 062308
\bibitem{cerf01}
Cerf NJ, Levy M and Van Assche G 2001 \textit{Phys. Rev. A} \textbf{63} 052311
\bibitem{grosshans03}
 Grosshans F, Van Assche G, Wenger J, Brouri R, Cerf N J and Grangier P 2003 \textit{Nature} \textbf{421} 6920
\bibitem{heine11}
 Heine F, Kampfner H, Greulich P, Seel S  2011 \textit{International Conference on Space Optical Systems and Applications (ICSOS)} pp. 286-289
\bibitem{lorenz04}
 Lorenz S, Korolkova N, Leuchs G 2004 \textit{Appl. Phys. B} \textbf{79} 273
\bibitem{lorenz06}
 Lorenz S, Rigas J, Heid M, Andersen U L, L\"{u}tkenhaus N and Leuchs G 2006 \textit{Phys. Rev. A} \textbf{74} 042326
\bibitem{elser09}
 Elser D, Bartley T, Heim B, Wittmann C, Sych D,  Leuchs G 2009 and \textit{New J. Phys.} \textbf{11} 045014
\bibitem{heim10}
 Heim B, Elser D, Bartley T, Sabuncu M, Wittmann C, Sych D, Marquardt Ch and Leuchs G 2010 \textit{Appl. Phys. B} \textbf{98} 4 635
\bibitem{lance05}
 A M Lance, T Symul, V Sharma, C Weedbrook, T C Ralph  and P K Lam  2005 \textit{Phys. Rev. Lett.} \textbf{95} 180503
\bibitem{sych10}
 Sych D and Leuchs G 2010 \textit{Optics and spectroscopy} \textbf{108} 3 326-330
\bibitem{wittmann10}
 Wittmann C, F\"{u}rst J, Wiechers C, Elser D, H\"{a}seler H, L\"{u}tkenhaus N  and G Leuchs 2010 \textit{Opt. Express}  \textbf{18} 5 4499
\bibitem{rigas06}
 Rigas J, G\"{u}hne O and L\"{u}tkenhaus N 2006 Phys. Rev. A \textbf{73} 012341
\bibitem{haeseler08}
 H\"{a}seler H, Moroder T and L\"{u}tkenhaus N 2008  Phys. Rev. A \textbf{77} 032303
\bibitem{killoran10}  Killoran N, H\"{a}seler H and L\"{u}tkenhaus N 2010 Phys. Rev. A \textbf{82} 052331 
\bibitem{killoran11}
 Killoran N and L\"{u}tkenhaus N 2011 \textit{Phys. Rev. A} \textbf{83} 052320
\bibitem{khan13}
  Khan I, Wittmann C, Jain N, Killoran N, L\"{u}tkenhaus N, Marquardt Ch and Leuchs G  2013 \textit{Phys. Rev. A}  \textbf{88} 010302(R)
\bibitem{semenov09}
 Semenov A A and Vogel W 2009 \textit{Phys. Rev. A} {\bf 80} 021802
\bibitem{usenko12}
 Usenko V C, Heim B, Peuntinger C, Wittmann C, Marquardt Ch, Leuchs G and Filip R 2012 \textit{New J. Phys.} \textbf{14} 093048
\bibitem{semenov12}
 Semenov A A, T\"{o}ppel F, Vasylyev D Yu, Gomonay H V and Vogel W 2012 \textit{Phys. Rev. A}  \textbf{85} 013826
\bibitem{peuntinger14}
 Peuntinger C, Heim B, M\"{u}ller C R, Gabriel C, Marquardt Ch and Leuchs G 2014 arXiv:1402.6290v1 [quant-ph]
\bibitem{heersink06}
 Heersink J, Marquardt Ch, Dong R, Filip R, Lorenz S, Leuchs G and Andersen U L 2006 \textit{Phys. Rev. Lett.} \textbf{96} 253601
\bibitem{dong08}
 Dong R, Lassen M, Heersink J, Marquardt Ch, Filip R, Leuchs G and Andersen U L 2008 \textit{Nat. Phys.} \textbf{4} 919
\bibitem {schnabel08}
 Hage B, Samblowski A, DiGuglielmo J, Franzen A, Fiur\'{a}\v{s}ek J and Schnabel R 2008 \textit{Nat. Phys} \textbf{4} 915
\bibitem{wittmann08}
 Wittmann C, Elser D, Andersen U L, Filip R, Marek P and Leuchs G 2008 \textit{Phys. Rev. A}  \textbf{78} 032315
\bibitem{erven12}
 Erven C, Heim B, Meyer-Scott E, Bourgoin J P, Laflamme R, Weihs G and Jennewein T 2012 \textit{New J. Phys.} \textbf{14}  123018
\bibitem{capraro12}
 Capraro I, Tomaello A, Dall'Arche A, Gerlin F, Ursin R, Vallone G and Villoresi P 2012 \textit{\PRL}  \textbf{109}  200502
\bibitem{stokes1852}
 Stokes G G 1852 \textit{Trans. Cambridge Philos. Soc.} \textbf{9} 399
\bibitem{korolkova02}
 Korolkova N, Leuchs G, Loudon R, Ralph T C and Silberhorn C 2002 \textit{Phys. Rev. A}  \textbf{65}  052306
\bibitem{stenholm92}
 Stenholm S 1992 \textit{Ann. Phys.} \textbf{218}  233
\bibitem{curty04}
 Curty M, Lewenstein M and L\"utkenhaus N 2004 \PRL \textbf{92} 217903
\bibitem{zhao09}
Zhao Y-B, Heid M, Rigas J and L\"utkenhaus N 2009 \textit{Phys. Rev. A}  \textbf{79}  012307
\bibitem{leverrier09er11}
Leverrier~A and Grangier~P 2009 \textit{\PRL}\textbf{102} 180504 and 
Leverrier~A and Grangier~P 2011 \textit{\PRL} \textbf{106} 259902
\bibitem{leverrier11}
Leverrier~A and Grangier~P 2011 \textit{Phys. Rev. A} \textbf{83} 042312
\bibitem{zyczkowsky98}
 Zyczkowski K,  Horodecki P, Sanpera A and Lewenstein M 1998 \textit{Phys. Rev. A}  \textbf{58} 883
\bibitem{lee00}
 Lee J, Kim M S, Park Y J and Lee S 2000 \textit{J. Mod. Opt.} \textbf{47} 2151
\bibitem{vidal02}
 Vidal G and Werner R F 2002 \textit{Phys. Rev. A}  \textbf{65}  032314
\bibitem{mosekyalmip} http://mosek.com; YALMIP: A Toolbox for Modeling and Optimization in MATLAB. J L\"ofberg. In Proceedings of the CACSD Conference, Taipei, Taiwan, 2004
\bibitem{plenio05}
 Plenio M B 2005 \textit{\PRL} \textbf{95}  090503
\bibitem{madsen12}
Madsen~L~R, Usenko~V~C, Lassen~M, Filip~R, Andersen~U~L 2012 \textit{Nat. Comm.} \textbf{4} 1083
\end{thebibliography}

\end{document}